\documentclass[aps,groupedaddress,superscriptaddress,showpacs]{revtex4-1}

\usepackage{graphicx}
\usepackage{color}
\usepackage{amsmath}
\usepackage{amsfonts}
\usepackage{amssymb}
\usepackage{dcolumn}
\usepackage{hyperref}
\usepackage{enumerate}
\usepackage{cancel}
\usepackage{mathtools}
\usepackage{multirow} 
\usepackage{appendix}
\usepackage{booktabs}
\usepackage{graphics}
\usepackage{graphicx}
\usepackage{changes}
\usepackage{float}
\restylefloat{table}

\begin{document}

\title{Random-Graph Models and Characterization of Granular Networks}
\author{Silvia Nauer}
\affiliation{Institute for Theoretical Physics, ETH Zurich, 8093, Zurich, Switzerland} 
\author{Lucas B\"ottcher}
\email{lucasb@ethz.ch}
\affiliation{Institute for Theoretical Physics, ETH Zurich, 8093, Zurich, Switzerland} 
\affiliation{Center of Economic Research, ETH Zurich, 8092, Zurich, Switzerland}
\author{Mason A. Porter}
\email{mason@math.ucla.edu}
\affiliation{Department of Mathematics, University of California, Los Angeles, CA 90095, USA}
\date{\today}
\begin{abstract}
Various approaches and measures from network analysis have been applied to granular and particulate networks to gain insights into their structural, transport, failure-propagation and other systems-level properties.
In this article, we examine a variety of common network measures and study their ability to characterize various two-dimensional and three-dimensional spatial random-graph models and empirical two-dimensional
granular networks. We identify network measures that are able to distinguish between physically plausible
and unphysical spatial network models. Our results also suggest that there are significant differences in the
distributions of certain network measures in two and three dimensions, hinting at important differences
that we also expect to arise in experimental granular networks.
\end{abstract}
\maketitle
%
%
%



\section{Introduction}
Various tools from network analysis have yielded insights into transport, failure mechanisms and other
system-level properties of granular networks, which are typically constructed by interpreting the particles of the underlying packings as nodes and interpreting their contacts as edges~\cite{smart2007effects,smart2008granular,berthier2018forecasting,papadopoulos2018network}. Network analysis may help lead to better understanding and the ability to control fracture processes, design new materials, and assess structural degradation in various engineering applications~\cite{herrmann2014statistical,anderson2017fracture,heisser2018controlling}. Researchers have employed continuum \cite{anderson2017fracture} and particle-level~\cite{herrmann2014statistical,cates1999jamming} descriptions to study granular materials \cite{jaeger1996,andreotti2013}, but neither framework is designed to study intermediate-scale organization, which is important for understanding and characterizing granular packings~\cite{bassett2015extraction,papadopoulos2018network}.

In studies of granular and particulate networks, it is not clear which network measures are most
suitable for distinguishing between physical granular networks and randomly embedded and overlapping
particle systems. To take advantage of methods of network analysis, it is necessary to identify appropriate
network measures to best highlight the distinctions between different structures~\cite{barthelemy2018complex}. Building on ideas from Ref.~\cite{porterstudentthesis}, we compute a variety of common network diagnostics and apply them to several spatial network models and empirical granular networks in two and three dimensions. We identify measures that
are able to distinguish between the network characteristics of physical and unphysical spatial networks
that we construct from random particle systems. We test the convergence properties of our results by
considering large ensembles of granular networks. Our results are useful both for the study of granular networks and because they help provide a roadmap for the development of random-graph models for
studying spatially-embedded networks (such as transportation, communication and vascular systems). As has also been demonstrated in applications like sensor systems \cite{coon2012impact,dettmann2016}, incorporating a small amount of physics into models such as random geometric graphs can provide important insights into the structure of such systems \cite{barthelemy2018complex}.

Our paper proceeds as follows. In Sec.~\ref{sec:network_diagnostics}, we describe the network measures that we use in our investigation of spatial random graphs and granular networks. In Sec.~\ref{sec:granular_models}, we present several models of both unphysical and physical granular networks. The network structures range from random geometric graphs (RGGs) to physically plausible contact networks. To ensure comparability of our results with properties of real granular networks, we also apply our network measures to empirical granular network structures, and we generate our simulated networks according to empirically observed edge densities. We thereby examine if certain network measures are able to distinguish between clearly unphysical (i.e., overlapping) granular packings and ones that are potentially physically plausible. In Sec.~\ref{sec:results}, we compare the distributions of the network diagnostics on the various networks. We also examine the convergence properties of the considered network diagnostics by computing them for ensembles of large networks. In Sec.~\ref{sec:discussion}, we conclude our study and discuss our results.
%
%
%


\section{Network diagnostics}
\label{sec:network_diagnostics}
The structure of a network depends fundamentally on the processes and interactions that define it~\cite{newman2018networks}. Many diagnostics have been developed to study and characterize networks, and we seek to use some of them to study granular networks and examine connections between network and particulate systems.

Consider a graph (i.e., network) $G(V,E)$, where $V$ is a set of nodes and $E$ is a set of edges. For each node $v \in V$, the degree $\text{deg}(v)$ is the number of edges that are attached to $v$. For each node pair $\langle s, t \rangle \in V$, let $d_{\text{min}}(s,t)$ denote a shortest path between $s$ and $t$. We use $\sigma_{st}$ to denote the number of shortest paths in the graph from $s$ to $t$. The number of shortest paths from $s$ to $t$ that traverse $v\in V$ is $\sigma_{st}(v)$, and $\sigma_{st}(e)$ denotes the number of shortest paths from $s$ to $t$ that traverse an edge $e\in E$. We let $T(G)$ denote the number of triangles (i.e., a set of three nodes with an edge between each pair of nodes) present in the graph. A triangle is a set of three nodes that are all pairwise connected by an edge. We denote the number of triangles that include a node $v \in V$ by $T(G;v)$. We count each triangle three times when computing the number of triangles in a graph $G$. All of the networks that we consider are unweighted and undirected, and we discuss several ways for characterizing them in the following paragraphs. To compute the values of these diagnostics, we use the Python library {\sc NetworkX}. See Ref.~\cite{networkx_doc} for a detailed documentation.
\paragraph{Edge density.}
The edge density (or sometimes just ``density'') of an undirected network is~\cite{newman2018networks}
\begin{equation}
	\rho = \frac{|E|}{|V|(|V|-1)}\,,
\end{equation}
where $|E|$ is the numbers of edges and $|V|$ is the number of nodes. We use edge density as a reference
measure for the construction of our networks. To compare the results of different network models, we
first make sure that their edge densities are the same. From this density and the number of nodes, we
derive the number of edges, which we use in our generation of random graphs.
\paragraph{Geodesic node betweenness.}
The geodesic node betweenness centrality $c_{B}$ for a node $v$ is the sum (over all node pairs $\langle s,t \rangle$ in a network) of the number of shortest paths from $s$ to $t$ that traverse $v$ divided by the total number of shortest paths from $s$ to $t$~\cite{newman2018networks}. Including a normalization factor, we write geodesic node betweenness centrality as
\begin{equation}
	c_{B}(v) = \sum_{\langle s,t \rangle \in V} \frac{\sigma_{st}(v)}{\sigma_{st}} \underbrace{\frac{2}{(|V| - 1)(|V| - 2)}}_{\text{normalization factor}}\,.
\end{equation}
Note that $\sigma_{st}=1$ if $s=t$ and that $\sigma_{st}(v)=0$ if $v\in\{s,t\}$.
We define a network's geodesic node betweenness
\begin{equation}
	C_B(G) =\frac{1}{|V|} \sum_{v \in V} c_{B}(v)
\end{equation}
as the mean of geodesic node betweenness centrality over all nodes in a network.

A large geodesic betweenness centrality $c_{B}(v)$ suggests that many shortest paths traverse a particular node. This may in turn suggest that such a node is relevant for transportation processes or failure-propagation mechanisms in granular networks~\cite{smart2007effects,smart2008granular,berthier2018forecasting,papadopoulos2018network}. 
\paragraph{Geodesic edge betweenness.}
We can also calculate a geodesic betweenness centrality for edges~\cite{girvan2002community}:
\begin{equation}
	c_{B}(e) = \sum_{\langle s,t \rangle \in V} \frac{\sigma_{st}(e)}{\sigma_{st}}\underbrace{\frac{2}{|V|(|V| - 1)}}_{\text{normalization factor}} \,.
\label{eq:geod_edge_betw_single}
\end{equation}
Note that $\sigma_{st} (e) = 0$ if $s = t$, and we again take $\sigma_{st} = 1$ when $s = t$. Based on Eq.~\eqref{eq:geod_edge_betw_single}, the geodesic edge betweenness of a network is
\begin{equation}
	C_B(G) = \frac{1}{|E|} \sum_{e \in E} c_{B}(e) \,.
\end{equation}
\paragraph{Clustering coefficient.}
The local clustering coefficient $c(v)$ of a node $v$ with degree $\text{deg}(v)\geq 2$ is the number of triangles that include that node divided by all possible triangles that could include $v$ (i.e., the connected triples)~\cite{newman2018networks}. That is,
\begin{equation}
	c(v) = \frac{2T(G;v)}{\text{deg}(v)(\text{deg}(v)-1)}\,.
\end{equation}
If the degree of a node is $0$ or $1$, the corresponding clustering coefficient of that node is $0$. The mean local clustering coefficient of a network is
\begin{equation}
	c(G) = \frac{1}{|V|}\sum_{v \in V} c(v) \, .
\end{equation}
\paragraph{Transitivity.}
Transitivity is the number of triangles in a network divided by the number of possible triangles~\cite{newman2018networks}:
\begin{equation}
	\text{Transitivity} = \frac{T(G)}{\text{Number of possible triangles in }G}\,.
\end{equation} 
Based on the definition of clustering coefficient and transitivity, we see that the two measures are closely related.
\paragraph{Degree assortativity.}
Degree assortativity is a measure to quantify the correlation between the degree of a certain node and the degrees of its neighbors~\cite{newman2018networks}. Larger values of degree assortativity entail greater similarity in the degree of neighboring nodes. We compute degree assortativity by calculating a Pearson correlation coefficient, whose values lie in the interval $[-1,1]$~\cite{Newman2003mixing}. A value of $1$ indicates a network with perfect degree assortativity, a value of $0$ corresponds to a nonassortative network, and a value of $-1$ indicates perfect degree disassortativity.
\paragraph{Global efficiency.}
The efficiency $E(u,v)$ of a node pair $\langle u,v \rangle$ is the inverse of the shortest-path length between these two nodes~\cite{latora2001efficient}. That is,
\begin{equation}
	E(u,v) = \frac{1}{d_{\text{min}}(u,v)}\,.
\end{equation}
Global efficiency is the mean of the efficiencies over all pairs of nodes in a network:
\begin{equation}
	E_{\text{glob}}(G) = \frac{1}{| \{ \langle u,v \rangle \in V\} |}\sum_{\langle u,v \rangle \in V} E(u,v)\,.
\end{equation}
\paragraph{Local efficiency.}
The local efficiency~\cite{latora2001efficient} of a node $v$ is
\begin{equation}
	E_{\text{loc}} (v) = \frac{1}{|{\langle u,w \rangle \in \Gamma(v) }|} \sum_{\langle u,w \rangle \in \Gamma(v)} E(u,w)\,,
\end{equation} 
where $\Gamma(v)$ denotes the neighborhood of $v$. The local efficiency of a network is
\begin{equation}
	E_{\text{loc}}(G) = \frac{1}{|V|} \sum_{v \in V} E_{\text{loc}}(v)\,,
\end{equation}
the mean of the local efficiencies over all nodes of the network.
\paragraph{Mean shortest-path length in the largest connected component (LCC).}
The mean shortest-path length over all node pairs $\langle u,v \rangle$ of the largest connected component (LCC) $C_{\text{max}}$ of a network is 
\begin{equation}
	\bar{d}_{\text{min}} = \frac{1}{| \{ \langle u,v \rangle \in C_{\text{max}}\} |}\sum_{\langle u,v \rangle \in C_{\text{max}}} d_{\text{min}}(u,v)\,.
\end{equation}
\paragraph{Weighted mean shortest-path length.}
The weighted mean shortest-path length $\bar{d}_{\text{min}}^w$ is the weighted arithmetic mean of the mean shortest-path lengths $\bar{d}_{\text{min}}(G_C)$ that we compute for connected components $G_C$ of a graph. We use the number of nodes in each component as a weight, so
\begin{equation}
	\bar{d}_{\text{min}}^w= \frac{1}{|V|} \sum_{G_C \in G} \bar{d}_{\text{min}}(G_C)\, |G_C|\,,
\end{equation}
where $|G_C|$ denotes the number of nodes in the considered connected component.
\paragraph{Maximized modularity.}
Subgraphs of a network $G$ are often called ``communities'' in the context of dense sets of nodes that are connected sparsely to other dense sets of nodes \cite{porter2009,fortunato2016}. Given a set of communities of a network $G$, the modularity $Q \in [-1, 1]$ measures the density of edges (adjusted for edge weights) inside communities
compared with that between communities~\cite{Blondel2008communities}. For an unweighted graph, the modularity $Q$ of a partition of a graph $G$ is
\begin{equation}
	Q = \frac{1}{2|E|} \sum_{\langle u,v \rangle \in G} \left[ A_{uv} - \frac{\text{deg}(u)\text{deg}(v)}{2|E|} \right] \delta(g_{u}, g_{v})\,,
\end{equation}
where $A$ is the adjacency matrix of the network, $\delta$ is the Kronecker delta, and $g_{u}$ denotes the community that includes node $u$.

We maximize modularity by applying the locally greedy Louvain method for community detection~\cite{Blondel2008communities}. Maximized modularity measures how well one can partition a network into disjoint
communities~\cite{porter2009}.
\paragraph{Subgraph centrality.}
For a node $v$ and an adjacency matrix $A$ associated with a graph $G$, the subgraph centrality~\cite{EstradaRodriguez}
\begin{equation}
	S(v) = \sum_{k=0}^{\infty} \frac{\left( A^{k} \right)_{vv}}{k!}
\end{equation}
is the sum of weighted closed walks (of all lengths) that start and end at node $v$. Instead of weighting each term with $1/k!$, one can also make other choices for the weights. The $(u,v)$th element of $A^{k}$ have the following interpretation: each element $(u,v)$ represents the number of paths of length $k$ from node $u$ to node $v$. 
The mean subgraph centrality of a network $G$ is
\begin{equation}
	S(G)=\frac{1}{|V|} \sum_{v \in V} S(v)\,.
\label{eq: subgraph_centrality}
\end{equation}
\paragraph{Communicability.}
The communicability~\cite{estrada2008communicability}
\begin{equation}
	C_{\mathrm{com}}(u, v) = \sum_{k = 0}^{\infty} \frac{(A^{k})_{uv}}{k!} \,,
\end{equation}
between nodes $u$ and $v$ measures the number of closed walks between nodes $u$ and $v$.

The communicability of a network $G$ is the mean of the communicabilities between all pairs of nodes in the network:
\begin{equation}
	C_{\mathrm{com}}(G) = \frac{1}{| \{ \langle u,v \rangle \in V \} |}\sum_{\langle u,v \rangle \in V} C_{\mathrm{com}}(u,v)\,.
\end{equation}
%
%
%
%
%
%


\section{Random-graph models of granular networks}
\label{sec:granular_models}
As baselines to compare to empirical granular networks, we consider various models of spatial networks
in two and three dimensions. We start with a standard RGG, which is an unphysical model, and then we
examine some more realistic models (such as force-modified RGGs). After discussing these models in two
dimensions, we also examine some models in three dimensions. The reason for studying different models,
ranging from unphysical ones to more physical ones, is that we aim to identify diagnostics that are able
to capture properties of realistic granular networks, which one constructs from non-overlapping particles
that interact with each other~\cite{porterstudentthesis}. The model networks also allow us to analyse the convergence properties
of network measures, because we can generate large numbers of different network configurations and
examine how they behave when they have progressively more nodes. We also apply our network measures
to two-dimensional empirical granular networks and use the empirically determined edge densities as
input for our granular network models in two dimensions. Given a specific edge density, we perform all
of the computations for the same configuration as in the experiments (see Sec.~\ref{subsec:Experimental_data}). Specifically, the empirical granular systems are confined by a two-dimensional box with dimensions $290~\text{mm} \times 380~\text{mm}$ and have a mean of $1122$ particles. To study models of three-dimensional materials, we examine model
networks with $1122$ particles in a box with dimensions $102~\text{mm} \times 102~\text{mm}\times 102~\text{mm}$.
%
%
%


\subsection{Random geometric graphs}
\begin{figure}
\centering
\includegraphics[width=\textwidth]{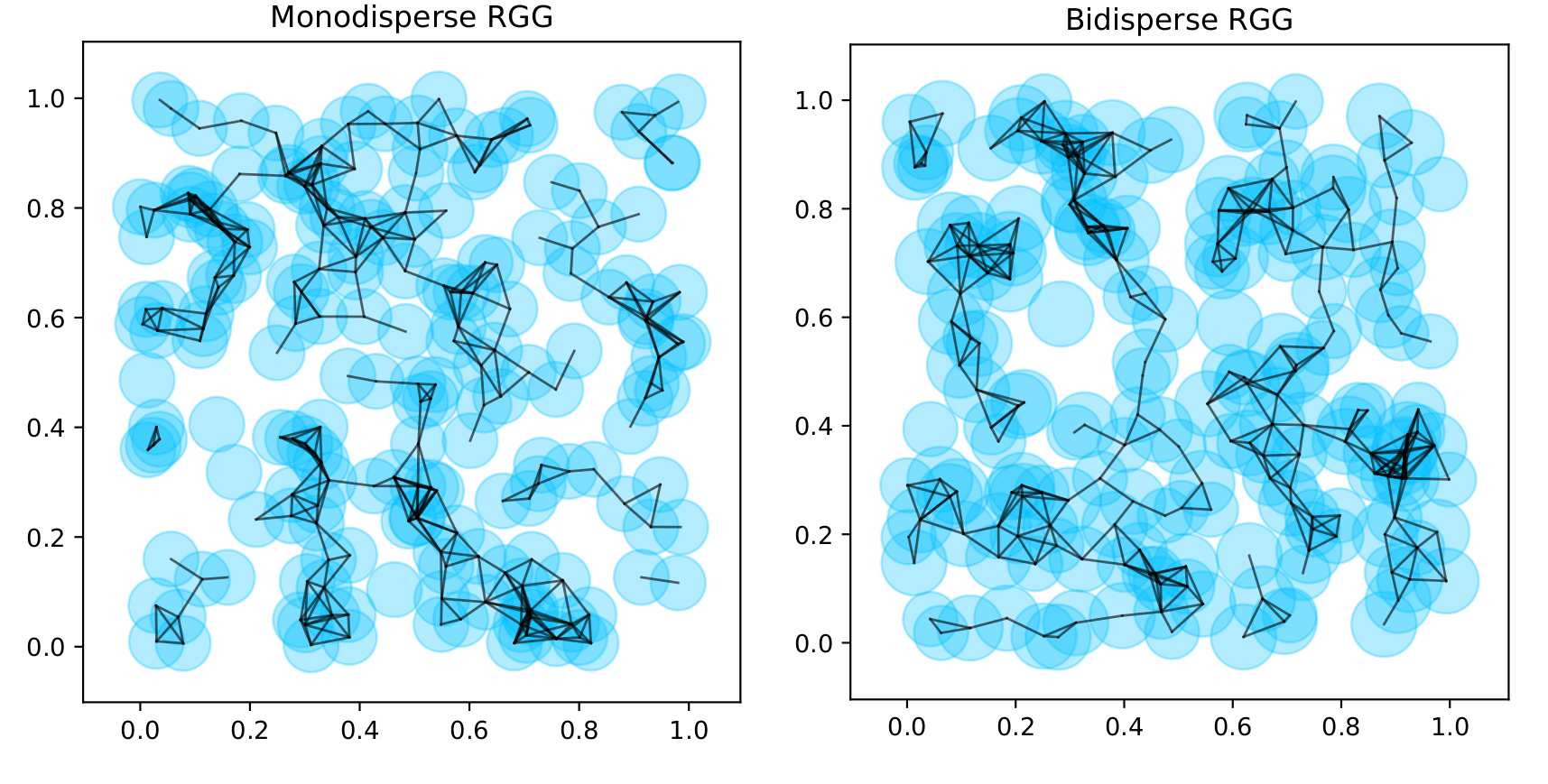}
\caption{\textbf{(Left) Monodisperse and (right) bidisperse RGG in the unit square.} The monodisperse RGG has $n=200$ nodes, particles of radius $r = 0.05$, and a distance parameter of $R \approx 0.0428$. The bidisperse RGG has $n = 200$ nodes, particles of radius $r_{1} = 0.05$ and $r_{2} = 0.06$ (there are $92$ of radius $r_1$ and $108$ particles of radius $r_2$), and a distance parameter of $R \approx 0.7770$. In the bidisperse case, we generate a uniformly distributed random number $\epsilon \sim \mathcal{U}(0,1)$ for each particle. If $\epsilon < 0.5$, we set the particle radius to $r_1$; otherwise, we set it to $r_2$.
} 
\label{fig:rgg}
\end{figure}
As a first, overly simplistic model of a granular system, we consider a monodisperse random geometric
graph (RGG). In this model, we place $n$ particles (represented by nodes) of the same radius $r$ uniformly at random in the box. The nodes are adjacent if the distance between them is smaller than $2R$, where we choose the parameter $R$ based on the edge density that matches an empirical granular network that we will study later (see Sec.~\ref{subsec:Experimental_data}). Because of the possibility of multiple overlapping particles, this monodisperse RGG is not a good model of a physical granular system. We also consider a bidisperse RGG. In this model, the particles can have two different radii, $r_{1}$ and $r_{2}$; nodes $i$ and $j$ are adjacent to each other if their distance is smaller than $R \left(r_{i} + r_{j}\right)$. For each particle, we generate a uniformly distributed random number $\epsilon \sim \mathcal{U}(0,1)$. If $\epsilon < 0.5$, we set the particle radius to $r_1$; otherwise, we set it to $r_2$. We use the same procedure for all bidisperse models in this article. We want 50\%, on average, of particles to have radius $r_1$ to facilitate our comparison with the employed experimental granular network data (see Sec.~\ref{subsec:Experimental_data}). As in the monodisperse case, we place particles uniformly at random. We again need to choose $R$ to match a desired edge density. We show examples of both a monodisperse and a bidisperse RGG in Fig.~\ref{fig:rgg}.

The parameters in our computations are $n = 1122$ nodes, a radius of $r = 4.5$, and a distance parameter of $R \approx 5.69$ for the monodisperse RGG and $n = 1122$, $r_{1} = 4.5$, $r_{2} = 5.5$, and $R \approx 1.13$ for the bidisperse RGG.
%
%
%


\subsection{Proximity-modified RGGs}
\begin{figure}
\centering
\includegraphics[width=\textwidth]{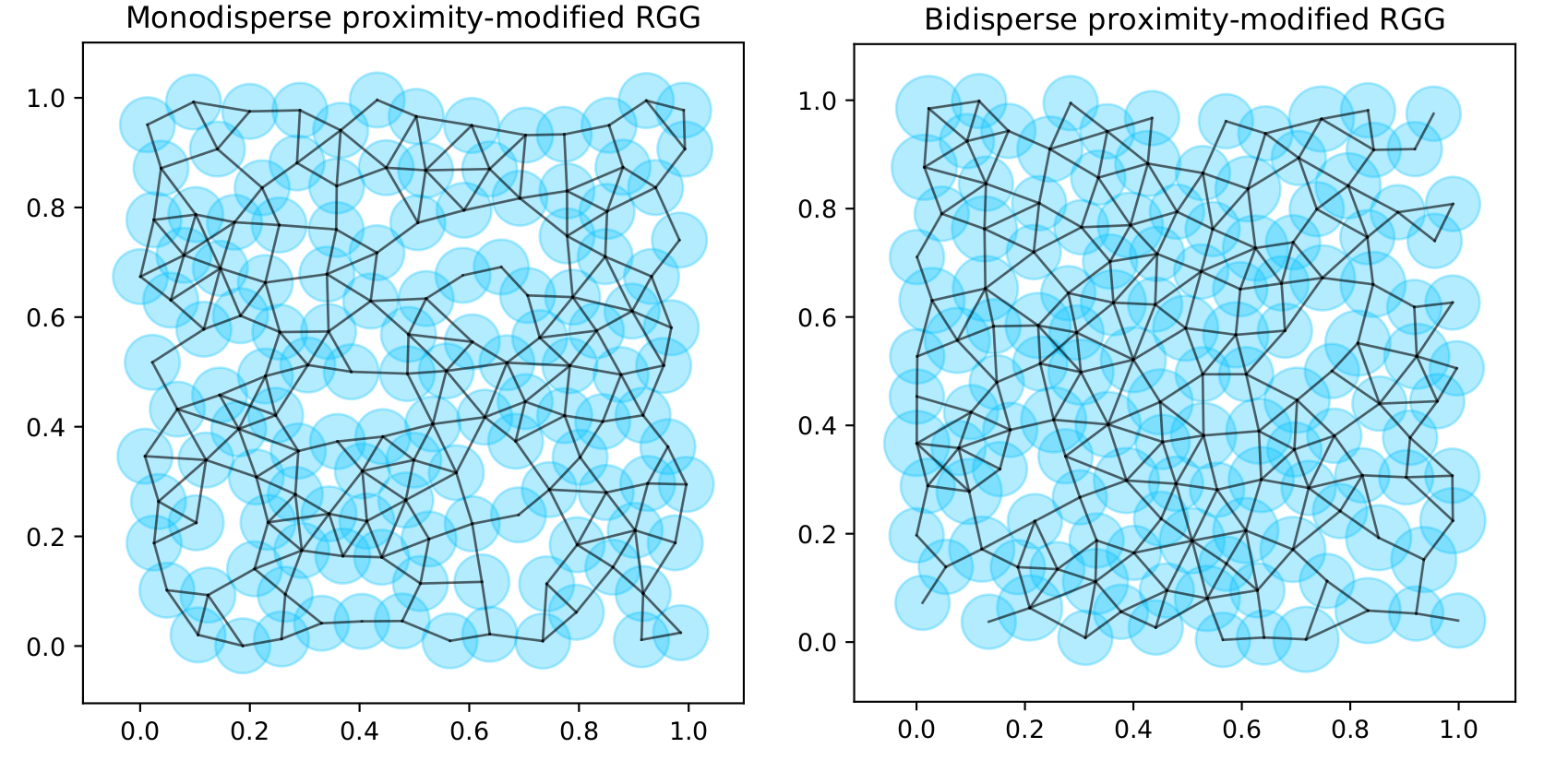}
\caption{\textbf{Monodisperse and bidisperse proximity-modified RGGs in the unit square.} The parameters are $n = 135$, $r = 0.05$, $\alpha = 0.7$, and $R \approx 0.0588$ for the monodisperse example and $n = 135$, $r_{1} = 0.05$, $r_{2} = 0.06$ (there are $90$ of radius $r_1$ and $45$ particles of radius $r_2$), $\alpha = 0.7$, and $R \approx 1.1020$ for the bidisperse example. The parameter $\alpha$ enforces a maximum particle density in a given region. In the bidisperse case, we generate a uniformly distributed random number $\epsilon \sim \mathcal{U}(0,1)$ for each particle. If $\epsilon < 0.5$, we set the particle radius to $r_1$; otherwise, we set it to $r_2$.}
\label{fig:proximity_modified}
\end{figure}
To compensate for some of the overlaps between particles in an RGG, we now define a proximity measure $p$ that describes the distance between each point $(x,y)$ and its nearest particle(s). Initially, the system is empty. We then place each particle as follows. For each integer $k \in \{1, \dots, n\}$, we place particle $k$, which has a radius of $r_{k}$, in the box. To do this, we choose a position $(x_{k}, y_{k})$ in the box uniformly at random, and we place the particle with probability
\begin{equation}\label{eq:1}
	P(p_{k}) = 
\begin{cases}
	0\,, & \text{for}\ p_{k} < 2\alpha r_{k}\,, \\
	1\,, & \text{for}\ p_{k} \geq 2\alpha r_{k}\,, 
\end{cases}
\end{equation}
where $p_{k}$ is the proximity measure for the position $(x_{k}, y_{k})$ and $\alpha \in[0,1]$ is a parameter that enforces a maximum particle density in a given region. We use this procedure for both monodisperse ($r_k=r$) and bidisperse ($r_k\in \{r_1,r_2\}$) particle configurations. We generate the edges in the same way as in the bidisperse RGG model. 

In Fig.~\ref{fig:proximity_modified}, we show examples of monodisperse and bidisperse proximity-modified RGGs. Based on the
depicted configurations, we see that the proximity-modified RGGs have fewer overlaps than the standard
RGGs. However, overlaps are still present, so the model does not describe a physical granular network.

For our computations, we use the parameters $n = 1122$, $r = 4.5$, $\alpha = 0.75$, and $R \approx 1.39$ for the monodisperse case and $n = 1122$, $r_{1} = 4.5$, $r_{2} = 5.5$, $\alpha = 0.75$, and $R \approx 1.24$ for the bidisperse case. In the bidisperse case, we generate a uniformly distributed random number $\epsilon \sim \mathcal{U}(0,1)$ for each particle. If $\epsilon < 0.5$, we set the particle radius to $r_1$; otherwise, we set it to $r_2$.
%
%
%


\subsection{Force-modified RGGs}
\label{subsec:force_mod_RGG}
\begin{figure}
\centering
\includegraphics[width=\textwidth]{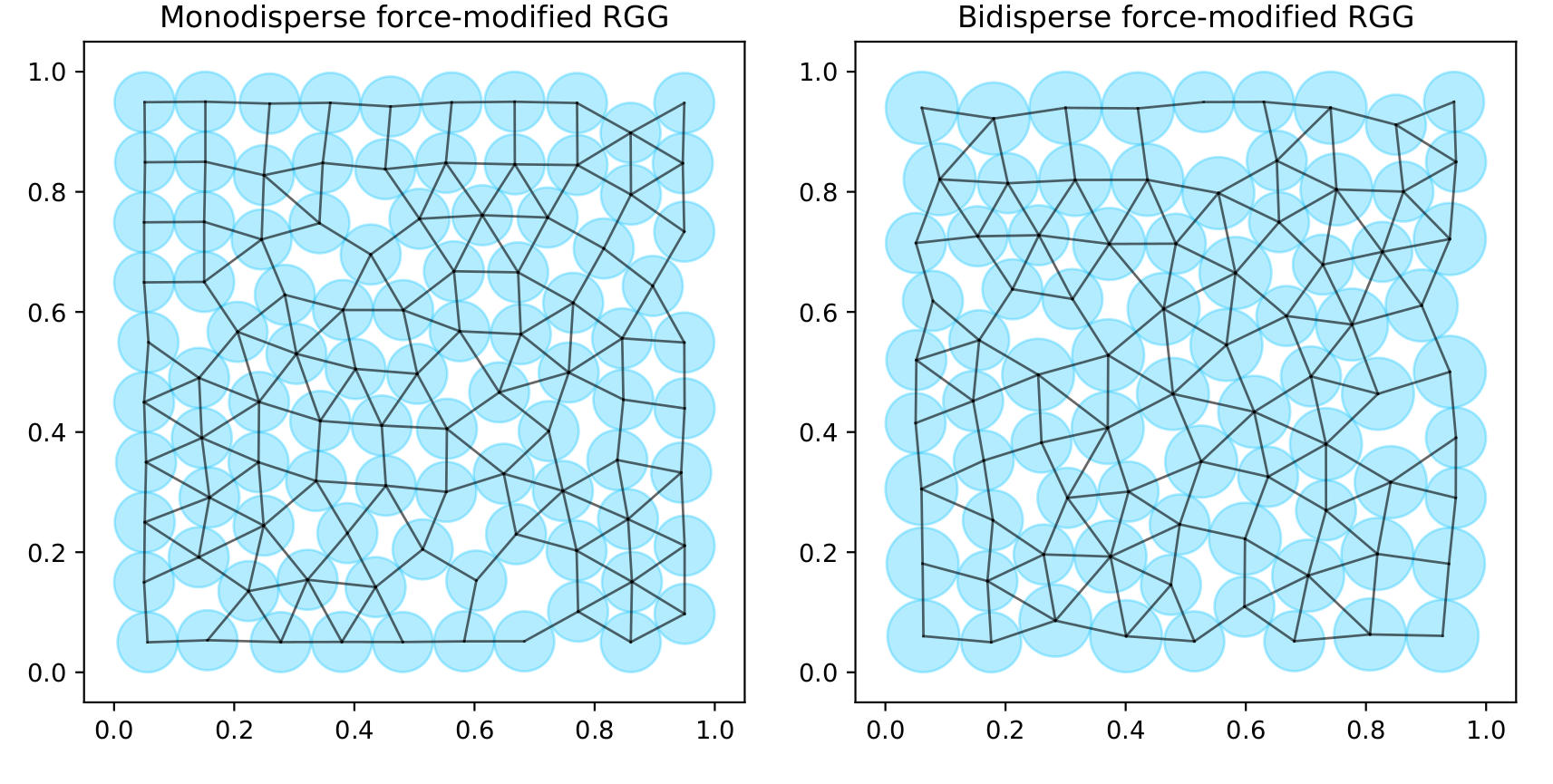}
\caption{\textbf{Monodisperse and bidisperse force-modified RGGs in the unit square.} The parameters are $n = 95$, $r = 0.05$, $\beta = 0.5$, and $R = 0.0615$ (monodisperse) and $n = 80$, $r_{1} = 0.05$, $r_{2} = 0.06$ (there are $41$ of radius $r_1$ and $39$ particles of radius $r_2$), $\beta = 0.5$, and $R \approx 1.2280$ (bidisperse). To generate these configurations, we update the positions ${\bf x}_k$ of all particles 700 times using ${\bf x}_k \rightarrow {\bf x}_k+{\bf d}_k$. The displacement ${\bf d}_k$ is proportional to the force ${\bf f}_k$ (see Eq.~\eqref{eq:force}), and the parameter $\beta$ gives the exponent in the force law. In the bidisperse case, we generate a uniformly distributed random number $\epsilon \sim \mathcal{U}(0,1)$ for each particle. If $\epsilon < 0.5$, we set the particle radius to $r_1$; otherwise, we set it to $r_2$.} 
\label{fig:force_modified}
\end{figure}
\begin{figure}
\centering
\includegraphics[width=\textwidth]{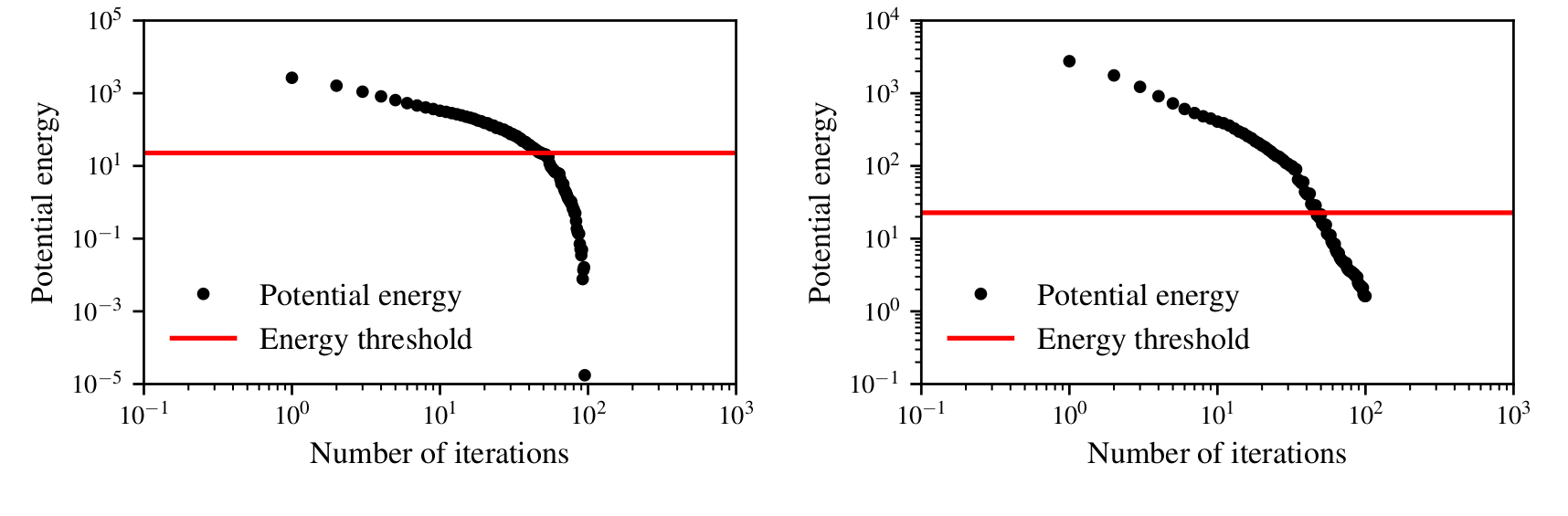}
\caption{\textbf{Energy convergence for the monodisperse force-modified RGG model.} We generate RGGs using the parameter values $n = 1122$, $r = 4.5$, $\beta = 0.5$, and $R \approx 1.29$ for a box with dimensions $L_{x} = 290$ and $L_{y} = 380$. For both of our examples, we update the positions ${\bf x}_k$ of all particles $100$ times using ${\bf x}_k \rightarrow {\bf x}_k+{\bf d}_k$. The displacement ${\bf d}_k$ is proportional to the force ${\bf f}_k$ defined by Eq.~\eqref{eq:force}. The left panel shows an example in which the energy approaches $0$, and the right panel shows an example in which the energy does
not approach 0 but is smaller than some threshold (which we indicate with the solid red line) that we choose such that there are no
visible overlaps between particles (see Appendix~\ref{sec:overlap}).
}
\label{fig:energy_fmrgg}
\end{figure}
To obtain a network whose characteristics better resemble those of a physical granular packing, we now
apply a different modification to an RGG. Initially, we start from an RGG configuration. Subsequently, for
each particle $k$, we suppose that particles interact via a Hertzian-like contact~\cite{owens2011sound} and compute the force $\mathbf{f}_{k}$ that acts on particle $k$ from overlaps with other particles. After computing $\mathbf{f}_{k}$ for all $k$, we update the
corresponding particle locations $\mathbf{x}_k$ according to $\mathbf{x}_k \rightarrow \mathbf{x}_k+\mathbf{d}_{k}$, with $\mathbf{d}_k=\varepsilon \mathbf{f}_{k}$. We repeat this process until the particles no longer overlap (see Appendix~\ref{sec:overlap}). One can use the parameter $\varepsilon$ to rescale the displacements $\mathbf{d}_k$, which may be too large and potentially lead to new overlaps between particles if we do not reduce
them. For a box with dimensions $L_{x}\times L_{y}$ and according to Hertzian contact theory, the force on particle $k$ is~\cite{owens2011sound,chong2017nonlinear} 
\begin{equation}
	\begin{aligned}
\mathbf{f}_{k} = & \sum_{l \neq k}\left\{ \left[\frac{1}{2}(r_{k} + r_{l}) - \frac{1}{2} |\mathbf{x}_{k} - \mathbf{x}_{l}|\right]_+^{\beta} \frac{\mathbf{x}_{k} - \mathbf{x}_{l}}{|\mathbf{x}_{k} - \mathbf{x}_{l}|}\right\} \\ 
& + \left[r_{k} - x_{k} \right]_+^{\beta}\hat{\mathbf{x}} - \left[r_{k} + x_{k} - L_{x} \right]_+^{\beta}\hat{\mathbf{x}} + \left[r_{k} - y_{k} \right]_+^{\beta}\hat{\mathbf{y}} - \left[r_{k} + y_{k} - L_{y} \right]_+^{\beta}\hat{\mathbf{y}}\,,
	\end{aligned}
\label{eq:force}
\end{equation}
where the exponent $\beta$ (which is $\beta = 3/2$ in classical Hertzian theory, but was estimated to be $\beta = 5/4$ for the 2D granular material in Ref.~\cite{owens2011sound}) describes the interaction strength between particles. The vectors $\hat{\mathbf{x}}$ and $\hat{\mathbf{y}}$ are the unit vectors in the $x$ and $y$ directions, respectively. Forces occur only between overlapping particles or between a particle and the box boundary if both are in contact. Therefore, the bracket $[x]_+$ in Eq.~\eqref{eq:force} is
\begin{equation}
	[x]_+=\begin{cases}
x\,, &\text{for } x>0\,,\\
0\,, &\text{for } x\leq0\,.
	\end{cases}
\end{equation}
\begin{figure}
\centering
\includegraphics[width=0.5\textwidth]{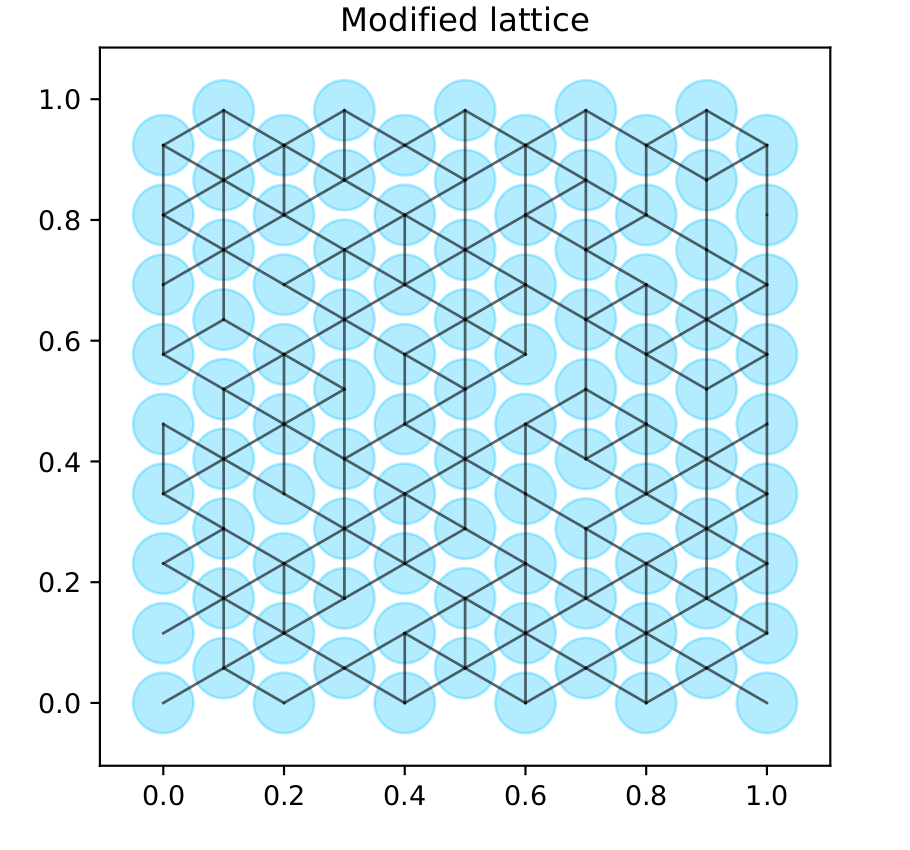}
\caption{\textbf{Modified lattice in the unit square.} We consider 99 particles of radius $r=0.05$ and a removal fraction of $\mu \approx 0.1870$.} 
\label{fig:modified_lattice}
\end{figure}
The potential energy that corresponds to Eq.~\eqref{eq:force} is
\begin{equation}
	\begin{aligned}
V = & \sum_{k}\sum_{l \neq k}\left\{ \left[\frac{1}{2}(r_{k} + r_{l}) - \frac{1}{2} |\mathbf{x}_{k} - \mathbf{x}_{l}|\right]_+^{\beta+1}\right\}
\\ & + \left[r_{k} - x_{k} \right]_+^{\beta+1} - \left[r_{k} + x_{k} - L_{x} \right]_+^{\beta+1} + \left[r_{k} - y_{k} \right]_+^{\beta+1} - \left[r_{k} + y_{k} - L_{y} \right]_+^{\beta+1}\,.
	\end{aligned}
\label{eq:potential}
\end{equation}
The first term of Eqs.~\eqref{eq:force} and \eqref{eq:potential} accounts for overlaps between particles, and the other terms take into
account the left, right, bottom and top boundaries of the box. We generate the edges in this model as we
did for bidisperse RGGs. We show examples of monodisperse and bidisperse force-modified RGGs in Fig.~\ref{fig:force_modified}. In this figure, it appears (based on visual inspection) that particles do not overlap anymore. At
least visually, they more closely resemble real granular networks than is the case for the standard RGGs
and the proximity-modified RGGs.

To attempt to ensure that our final configurations are free of overlaps, we monitor their energy.
If the energy is 0, there are no overlaps, so we stop updating locations when we cannot distinguish
the energy from 0. However, in some cases, the energy does not approach 0, because some particles
are locally jammed~\cite{behringer2018physics}. In such cases, we accept a configuration if its energy is smaller than the energy of a system in which every particle $k$ has an overlap distance of no more than $r_{k}/200$ with all of its neighbors. In Fig.~\ref{fig:energy_fmrgg}, we show the monitored energy for two cases of the monodisperse force-modified RGG. A situation in which many particles overlap with their neighbours is unlikely to occur.
Instead, there are often some particles that overlap with one or two neighbours with a large fraction of their areas. To avoid large overlaps, we also check for every configuration that each particle $k$ does not have an overlap distance with a neighbour that is larger than $r_{k}/15$. See Appendix~\ref{sec:overlap} for details.

The parameters in our computations are $n = 1122$, $r = 4.5$, $\beta = 0.5$, $L_{x} = 290$, $L_{y} = 380$, and $R \approx 1.2900$ for the monodisperse case. For the bidisperse case, they are $n = 1122$, $r_{1} = 4.5$, $r_{2}= 5.5$, $\beta=0.5$, $L_{x} = 290$, $L_{y} = 380$, and $R \approx 1.0850$. To generate these configurations, we perform $100$ iterations for the monodisperse case and $400$ iterations for the bidisperse one. As in the RGG model, 50\% of the particles, on average, in the bidisperse case have a radius of $r_1$; and the others have a radius of $r_2$.


\subsection{Modified lattice}

We also consider a modified lattice. In this model, we modify a hexagonal lattice of particles of the same
radii by removing a fraction $\mu$ of edges uniformly at random to match a certain edge density. We show an example of a modified lattice in Fig.~\ref{fig:modified_lattice}. In our computations, we use a total of 1020 particles with $r = 4.5$ and $\mu \approx 0.26$.
%
%
%


\subsection{Experimental data}
\label{subsec:Experimental_data}
\begin{figure}
\centering
\includegraphics[width=\textwidth]{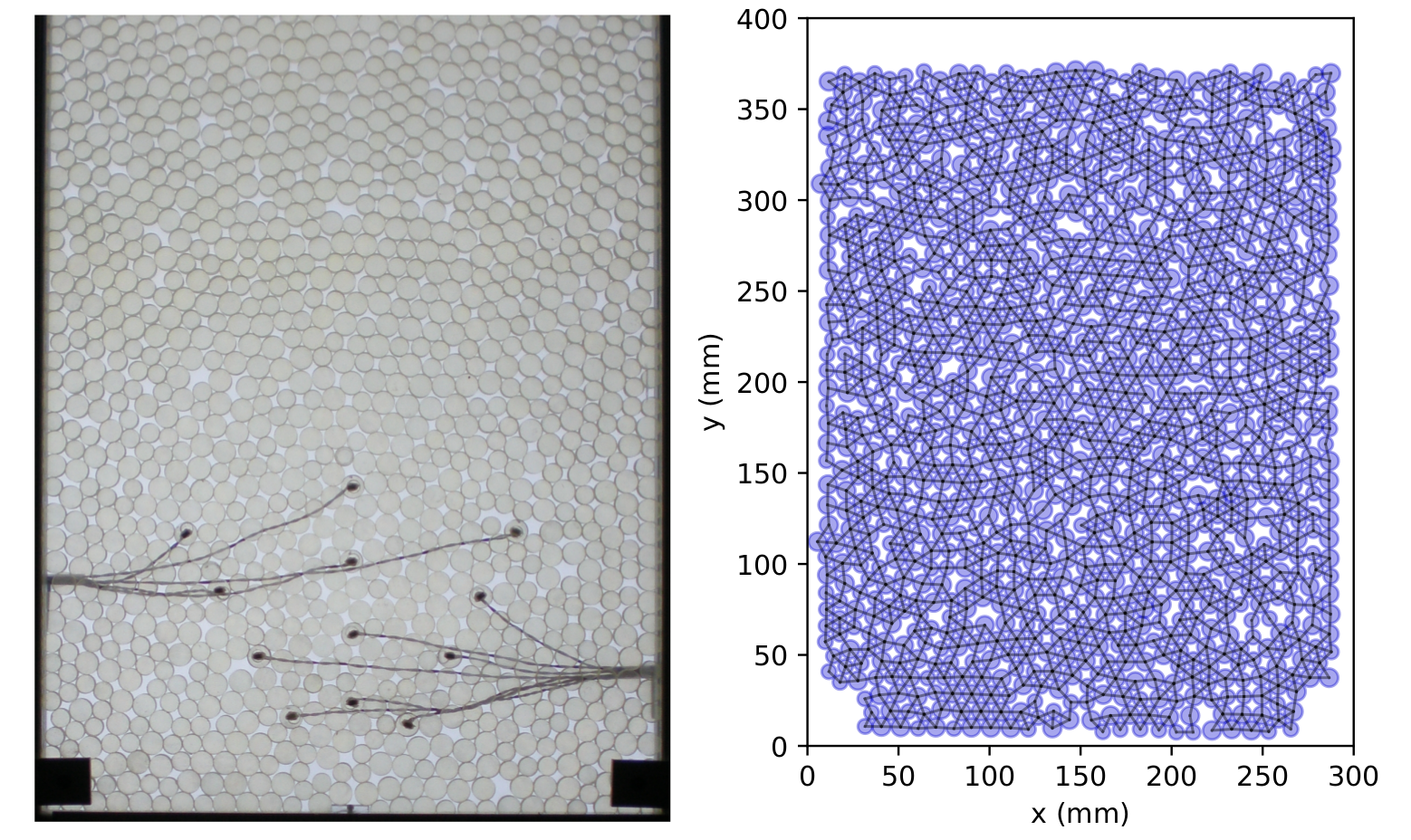}
\caption{\textbf{Experimental granular network.} The particles have radii $r_{1}=4.5~\text{mm}$ and $r_{2} = 5.5~\text{mm}$, and the box has sides of lengths $L_{1} = 290~\text{mm}$ and $L_{2} =380~\text{mm}$. In the left panel, we show the experimental setup; in the right panel, we show the corresponding extracted granular network. The data is taken from Ref.~\cite{bassett2012influence}.
} 
\label{fig:experimental_data}
\end{figure}
\begin{figure}[htp!]
\centering
\includegraphics[width=\textwidth]{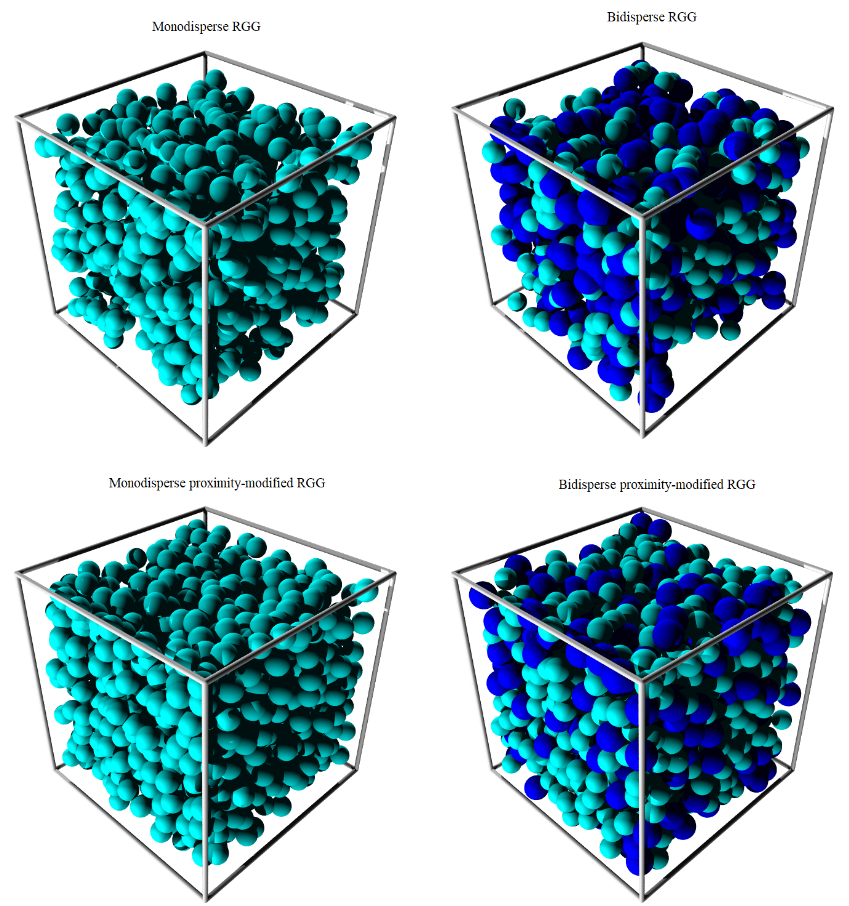}
\caption{\textbf{Three-dimensional RGG and proximity-modified RGG models.} For the standard RGGs, the parameters are $n = 1122$, $r = 4.5$, and $R \approx 6.0750$ for the monodisperse case and $n = 1122$, $r_{1} = 4.5$, $r_{2} = 5.5$, and $R \approx 1.1950$ for the bidisperse case. For the proximity-modified RGG models, the parameters are $n = 1122$, $r = 4.5$, $\alpha = 0.75$, and $R \approx 1.4200$ for the monodisperse case and $n = 1122$, $r_{1} = 4.5$, $r_{2} = 5.5$, $\alpha = 0.75$, and $R \approx 1.4385$ for the bidisperse cases. In the bidisperse case, we generate a uniformly distributed random number $\epsilon \sim \mathcal{U}(0,1)$ for each particle. If $\epsilon < 0.5$, we set the particle radius to $r_1$; otherwise, we set it to $r_2$. In all four examples, we consider a box of dimensions $L_x\times L_y \times L_z=102\times102\times 102$.}
\label{fig:3D_RGG_prox}
\end{figure}

To compare our model networks with real granular networks, we extract granular networks from experimental data~\cite{bassett2012influence}. Henceforth, when we write ``experimental data'', we mean data from a bidisperse granular system with radii $r_{1}=4.5~\text{mm}$ and $r_{2} = 5.5~\text{mm}$ in a two-dimensional box with side lengths $L_{1} =290~\text{mm}$ and $L_{2} =380~\text{mm}$. The employed experimental data set consists of $17$ different configurations of granular networks. We compute the diagnostics of Sec.~\ref{sec:network_diagnostics} for these $17$ cases, and we present the corresponding diagnostic values in Appendix~\ref{sec:diagnostic_values}. On average, the number of particles in the experimental configurations is $1122$, and $49.2$\% of the particles have a radius of $r_1$. Therefore, we use $1122$ particles in our model networks to facilitate comparisons between our models and the experimental data. For the bidisperse network models, we generate a uniformly distributed random number $\epsilon\sim\mathcal{U}(0,1)$. If $\epsilon < 0.5$, we set the particle radius to $r_1$; otherwise, we set it to $r_2$.
We consider bidisperse packings, as we wish to avoid the crystallization of monodisperse packings. We show an example of the experimental setup in Fig.~\ref{fig:experimental_data}. To
obtain such a configuration, Ref.~\cite{bassett2012influence} placed about $1000$ particles (which were cut from Vishay PSM-4 photoelastic material) into a container with an open top, such that the particles were confined only by
gravity.


\subsection{Models in 3D}
\label{subsec:3D_models}
\begin{figure}[htp!]
\centering
\includegraphics[width=\textwidth]{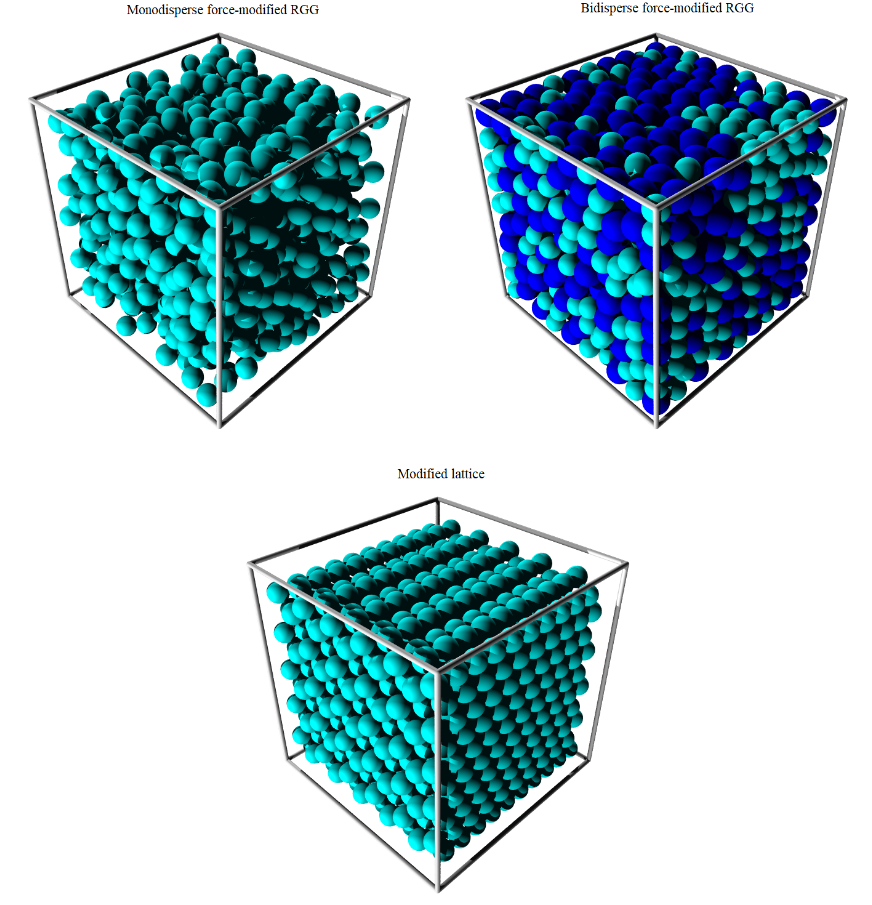}
\caption{\textbf{Three-dimensional force-modified RGG and modified lattice models.} For the force-modified RGG models, the parameters
are $n = 1122$, $r = 4.5$, $\beta=0.5$, and $R \approx 1.4080$ for the monodisperse case and $n = 1122$, $r_{1} = 4.5$, $r_{2} = 5.5$, $\beta=0.5$, and $R \approx 1.2600$ for the bidisperse case. To generate these configurations, we update the positions ${\bf x}_k$ of all particles $100$ times for the monodisperse case and 400 times for the bidisperse case according to ${\bf x}_k \rightarrow {\bf x}_k+{\bf d}_k$. The displacement ${\bf d}_k$ is proportional to the force ${\bf f}_k$ (see Eq.~\eqref{eq:force3D}). For the modified lattice, we arrange $1100$ particles of radius $r = 4.5$ in a hexagonal packing and we set $\mu \approx 0.4950$. In the bidisperse case, we generate a uniformly distributed random number $\epsilon \sim \mathcal{U}(0,1)$ for each particle. If $\epsilon < 0.5$, we set the particle radius to $r_1$; otherwise, we set it to $r_2$. In all cases, we consider a box of dimensions $L_x\times L_y \times L_z=102\times102\times 102$.}
\label{fig:3D_force_hex}
\end{figure}
As with the 2D examples, we calculate the network diagnostics of Sec.~\ref{sec:network_diagnostics} for 3D granular networks. In this section, we briefly discuss extensions of the previously-discussed
models to 3D. The standard RGGs do not require any special modification for the generation of the configurations, although the underlying box is now in 3D. We show examples of 3D monodisperse and bidisperse RGGs in the top panels of Fig.~\ref{fig:3D_RGG_prox}. As in the 2D RGGs, we observe that the particles overlap, so the model is not physical. Our parameters for these models are $n = 1122$, $r = 4.5$, and $R \approx 6.0750$ for the monodisperse case and $n = 1122$, $r_{1} = 4.5$, $r_{2} = 5.5$, and $R \approx 1.1950$ for the bidisperse case. As in the 2D models, 50\% of the particles in our 3D models have radius $r_1$ on average and the others have radius $r_2$.

It is also straightforward to extend the proximity-modified RGGs to 3D. We show
examples of 3D monodisperse and bidisperse proximity-modified RGGs in the bottom
panels of Fig.~\ref{fig:3D_RGG_prox}. We see that the particles overlap less than for the unmodified 3D RGGs, but they still overlap. To obtain these configurations, we use the parameters $n = 1122$, $r = 4.5$, $\alpha = 0.75$, and $R \approx 1.4200$ for the monodisperse case and $n = 1122$, $r_{1} = 4.5$, $r_{2} = 5.5$, $\alpha = 0.75$, and $R \approx 1.4385$ for the bidisperse case.

For the force-modified RGG, we have to adapt Eqs.~\eqref{eq:force} and \eqref{eq:potential} to 3D. Equation~\eqref{eq:force} becomes
\begin{equation}
	\begin{aligned}
\mathbf{f}_{k} = & \sum_{l \neq k}\left\{ \left[\frac{1}{2}(r_{k} + r_{l}) - \frac{1}{2} |\mathbf{x}_{k} - \mathbf{x}_{l}|\right]_+^{\beta} \frac{\mathbf{x}_{k} - \mathbf{x}_{l}}{|\mathbf{x}_{k} - \mathbf{x}_{l}|}\right\}
\\ & + \left[r_{k} - x_{k} \right]_+^{\beta}\hat{\mathbf{x}} - \left[r_{k} + x_{k} - L_{x} \right]_+^{\beta}\hat{\mathbf{x}} + \left[r_{k} - y_{k} \right]_+^{\beta}\hat{\mathbf{y}} - \left[r_{k} + y_{k} - L_{y} \right]_+^{\beta}\hat{\mathbf{y}}
\\ & + \left[r_{k} - z_{k} \right]_+^{\beta}\hat{\mathbf{z}} - \left[r_{k} + z_{k} - L_{z} \right]_+^{\beta}\hat{\mathbf{z}}\,,
	\end{aligned}
\label{eq:force3D}
\end{equation}
where $\hat{\mathbf{z}}$ denotes the unit vector in the $z$ direction. Equation~\eqref{eq:potential} becomes
\begin{equation}
	\begin{aligned}
V = & \sum_{k}\sum_{l \neq k}\left\{ \left[\frac{1}{2}(r_{k} + r_{l}) - \frac{1}{2} |\mathbf{x}_{k} - \mathbf{x}_{l}|\right]_+^{\beta+1}\right\}
\\ & + \left[r_{k} - x_{k} \right]_+^{\beta+1} - \left[r_{k} + x_{k} - L_{x} \right]_+^{\beta+1} + \left[r_{k} - y_{k} \right]_+^{\beta+1} - \left[r_{k} + y_{k} - L_{y} \right]_+^{\beta+1}
\\ & + \left[r_{k} - z_{k} \right]_+^{\beta+1} - \left[r_{k} + z_{k} - L_{z} \right]_+^{\beta+1}\,.
	\end{aligned}
\label{eq:potential3D}
\end{equation}

We show examples of 3D force-modified RGGs in the top panels of Fig.~\ref{fig:3D_force_hex}. As in the 2D case, we conclude based on visual inspection that they resemble real granular networks
more closely than the other model. The parameters in our computations are $n = 1122$, $r = 4.5$, $\beta=0.5$, and $R \approx 1.4080$ for the monodisperse case and $n = 1122$, $r_{1} = 4.5$, $r_{2} = 5.5$, $\beta=0.5$, and $R \approx 1.2600$ for the bidisperse case. In both examples, we set $L_x= L_y= L_z=102$. To generate these configurations, we perform $100$ iterations for the monodisperse case and $400$ iterations for the bidisperse one. We describe the energy thresholds for the 3D force-modified RGGs in Appendix~\ref{sec:overlap}.

We extend the modified lattice to 3D by considering spheres in a hexagonal close
packing (HCP). An HCP consists of a series of two alternating layers of spherical particles, where each
layer has a hexagonal arrangement. We create the edges in the same way as in the 2D
modified lattice. In the bottom panel of Fig.~\ref{fig:3D_force_hex}, we show an example of a 3D modified lattice that consists of $1100$ particles. The particles have a radius of $r = 4.5$, and we remove a fraction $\mu \approx 0.4950$ of the edges uniformly at random from the network.


\section{Results}
\label{sec:results}
%
%
%


\subsection{Diagnostics for 2D granular networks}
We first focus on the results of calculating the network diagnostics for the 2D models from Sec.~\ref{sec:granular_models}. We describe the convergence characteristics of all diagnostics and then discuss the distributions of several of them.


\subsubsection{Convergence characteristics}
\label{subsec:convergence}
\begin{figure}[htp!]
\centering
\includegraphics[width=\textwidth]{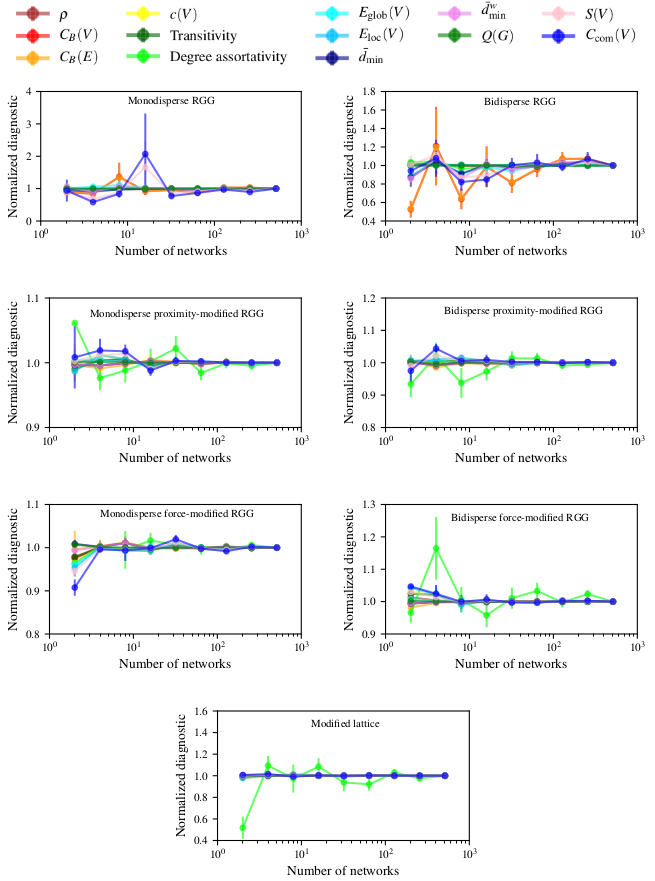}
\caption{\textbf{Convergence characteristics of the diagnostics.} We show the mean diagnostic values and the corresponding error bars for ensembles with different numbers of networks. We normalize each diagnostic value by the mean that we compute from the largest ensemble of networks.} 
\label{fig:diag_convergence}
\end{figure}

To monitor the convergence of the diagnostics, we generate $10$ ensembles with different numbers of granular-network configurations. Specifically, ensemble $m$ has $2^m$ networks (where $m \in \{0,\dots,9\}$), so there are $1023$ networks in total. For each $m$, we compute the mean of the diagnostic values and normalize each value by the mean that we calculate for $m=9$. Therefore, each of the mean diagnostic values approaches $1$ for $m=9$ (see Fig.~\ref{fig:diag_convergence}). This procedure allows us to identify a reasonable number of networks such that the deviations of the mean diagnostic values are sufficiently small in comparison to the values in the largest ensemble. We thereby study the convergence properties of the mean diagnostic values with respect to the values in the largest ensemble. Based on the data in Fig.~\ref{fig:diag_convergence}, we see that the diagnostics behave differently in the different network models.

In the unmodified RGG models, some of the normalized mean diagnostics deviate more from $1$ than in the other models. We observe this behavior for geodesic node betweenness, geodesic edge betweenness, subgraph centrality, and communicability. Because we place the particles in the RGGs uniformly at random, different configurations can differ substantially from each other. For example, in our computations, we observe that some particle configurations have local accumulations of particles, and such outliers strongly influence the mean over ensembles with a small number of networks. Such outliers exert less influence for ensembles with many networks, so they have a smaller effect on those ensembles. For the aforementioned diagnostics, such particle accumulations in some samples lead to the clearly visible zigzag behavior in Fig.~\ref{fig:diag_convergence}. In the other network models, this issue is less prominent, because local particle accumulations are suppressed by the proximity and force modifications. The larger deviations of the mean diagnostics in the unmodified RGG models is also reflected by their distributions (see Sec.~\ref{subsec:distribution}). For the proximity-modified RGGs and the force-modified RGGs, the mean diagnostic values approach
the values that we observe in the largest ensemble faster than in the unmodified RGG models. For the
modified-lattice model, we observe that almost all mean diagnostic curves are constant; the only exception
is degree assortativity. In this model, the only process that has any randomness is deleting edges from the
initially-complete hexagonal structure (in 2D) or HCP packing (in 3D).
%
%
%


\subsubsection{Numerical values}
\label{subsec: numerical_values}
\begin{figure}[htp!]
\centering
\includegraphics[width=\textwidth]{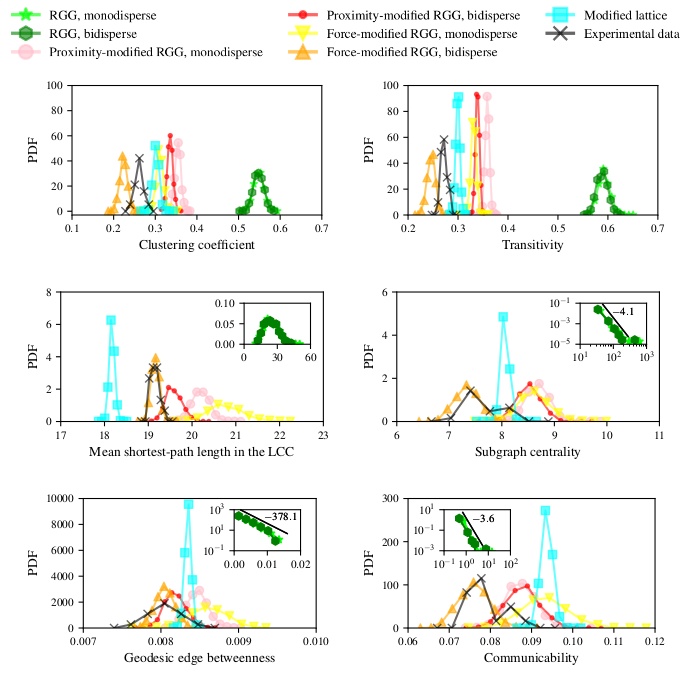}
\caption{\textbf{Distributions of the diagnostics in the 2D models and the experimental data.} We show the distributions of the
mean local clustering coefficient, transitivity, mean shortest-path length in the LCC, subgraph centrality, geodesic edge betweenness
and communicability for the 2D models and the experimental data. We compute these distributions from the 1023
network realizations for each model and the 17 realizations for the experimental data. The black lines in the insets are guides to
the eye that are based on power-law and exponential fits (using the method of least squares) of distributions for standard bidisperse
RGGs. (These insets also include the distributions for the standard monodisperse RGGs, but the fits use data only from the bidisperse
RGGs.) The corresponding numbers indicate the slopes of these lines.
} 
\label{fig:diag_distribution}
\end{figure} 

We summarize the diagnostic values for the models and experimental granular networks in Appendix~\ref{sec:diagnostic_values}. For each model, our results are means over all $1023$ configurations that we generated to study the convergence characteristics of the diagnostics in Sec.~\ref{subsec:convergence}. For the experimental data, we take means over the $17$ network realizations.

Based on the numbers in Tables~\ref{tab:diag_exp_hex_RGG} and \ref{tab:diag_prox_force} in Appendix~\ref{sec:diagnostic_values}, we note several differences between our models. From the results, it is clear that the standard RGGs are the most unphysical of these model; we already made this point when visualizing an associated particle system. The proximity-modified RGGs have diagnostic values that resemble the experimental values more closely than the standard RGGs, and the force-modified RGGs generate the most realistic configurations. The diagnostic values for the
modified-lattice model agree partly with those in the experimental data, but they are not better than the
values for the force-modified RGGs, and the values of degree assortativity in the modified-lattice model differ significantly from those in the experimental configurations. The modified lattices are also visually rather different from the experimental packings, as noted in Ref.~\cite{porterstudentthesis}. A closer look at the diagnostics of the bidisperse force-modified RGG reveals that it is the model that best matches the experimental data. This is also apparent from the distributions in Sec.~\ref{subsec:distribution} and Appendix~\ref{sec:distributions2}. 


\subsubsection{Distributions of network characteristics}
\label{subsec:distribution}
We now discuss the distributions of the network characteristics in ensembles of networks for the 2D models and the experimental data. We show the distributions of the clustering coefficient, transitivity, mean shortest-path length in the LCC, subgraph centrality, geodesic edge betweenness, and communicability in Fig.~\ref{fig:diag_distribution}. We show the distributions of the other measures in Appendix~\ref{sec:distributions2}. In each plot, the black curve is the distribution of the experimental data.

The top panels of Fig.~\ref{fig:diag_distribution} suggest that the mean local clustering coefficient and transitivity are useful
measures for distinguishing between the unphysical and physical models. Specifically, the distributions
of the standard RGGs do not overlap with the distributions from the other models, so we can clearly
distinguish between the RGG distributions and those of the other models. A closer look at the distributions
of mean local clustering coefficient and transitivity also reveals that the bidisperse force-modified RGG distributions and the empirical granular networks have characteristics that are distinct from those of the other models. Based on these results, it is possible to distinguish these two network types from the proximity-modified
RGGs and the monodisperse force-modified RGG. The modified-lattice distribution overlaps slightly
with the one from the experimental data, although visual inspection of the networks indicates that these
two types of networks are rather distinct from each other.

The distributions of the mean shortest-path length in the LCC in the experimental granular networks
are similar to those that we obtain for the bidisperse force-modified RGG. By contrast, the distributions
of the mean local clustering coefficient and transitivity in the experimental networks differ from those in all of the models. This difference is extremely large for the unphysical models, but there is some overlap
between experimental and model distributions for other models (e.g., more physical ones). Therefore,
computing these quantities may not yield a clear distinction between experimental and model distributions
when sampling from data.

In the remaining three panels of Fig.~\ref{fig:diag_distribution} we show the distributions of subgraph centrality, geodesic
edge betweenness and communicability. In the corresponding insets, we show the distributions for the
standard monodisperse and bidisperse RGG models, and we conclude that they differ qualitatively from
the distributions of the corresponding diagnostics in the other models. We can connect this observation
to the convergence characteristics of subgraph centrality, geodesic edge betweenness and communica-
bility. To study these convergence characteristics, we compare the mean values of all diagnostics for
different numbers of network realizations to the corresponding mean values in the largest ensemble (see Sec.~\ref{subsec:convergence}). For subgraph centrality, geodesic edge betweenness and communicability in the unmodi-
fied RGGs, we observe larger deviations of the mean diagnostic values in the smaller ensembles from the
mean values in the largest ensemble than is the case for the other diagnostics (see Fig.~\ref{fig:diag_convergence}). This observation
is also reflected by the qualitatively different distributions of those diagnostics (see Fig.~\ref{fig:diag_distribution}).

%
%
%


\subsection{Diagnostics for 3D granular networks}
\begin{figure}[htp!]
\centering
\includegraphics[width=\textwidth]{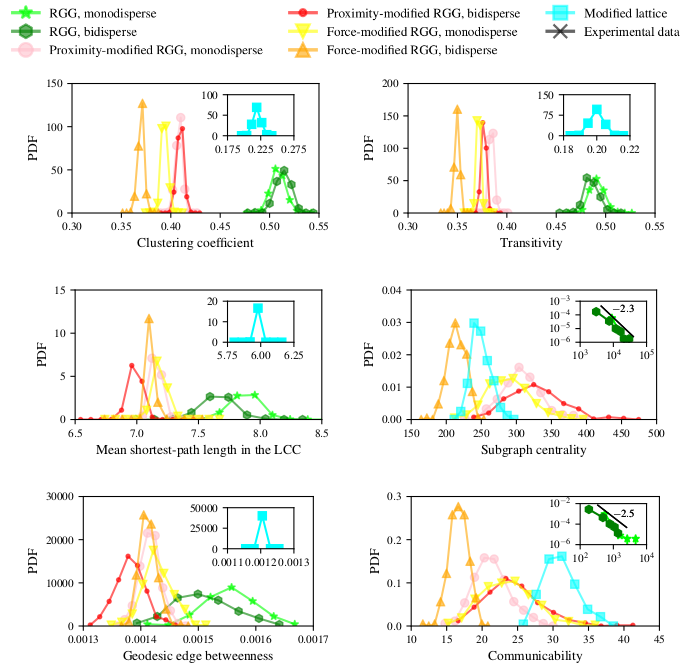}
\caption{\textbf{Distributions of the diagnostics in 3D.} We show the distributions of the mean local clustering
coefficient, transitivity, mean shortest-path length in the LCC, subgraph centrality, geodesic edge betweenness and communicability
for the three-dimensional network models. We base the distributions on 256 network realizations for each model. The black lines in
the insets are guides to the eye that we obtain from power-law fits (using the method of least squares) of the distributions for standard
bidisperse RGGs. (These insets also include the distributions for the standard monodisperse RGGs, but the fits use data only from
the bidisperse RGGs.) The corresponding numbers indicate the slopes of these lines. In some panels, we show the distribution for
the modified-lattice model in an inset.
}
\label{fig:3D_diag_distribution}
\end{figure} 
We now briefly discuss the behavior of the diagnostic distributions for the 3D network models. We again focus on mean local clustering coefficient, transitivity, mean shortest-path length in the
LCC, subgraph centrality, geodesic edge betweenness and communicability. We show the corresponding
distributions in Fig.~\ref{fig:3D_diag_distribution} and the remaining ones in Appendix~\ref{sec:distributions2}. For each model, we base the distributions on $256$ network realizations.

In the top panels of Fig.~\ref{fig:3D_diag_distribution}, we show the distributions of mean local clustering coefficient and
transitivity. As in our 2D models, these two diagnostics distinguish successfully between
physical and unphysical models of granular networks. The only major difference between our results
in 2D and 3D are the distributions (which we show in the insets in some panels) in the
modified-lattice model. In 3D, the mean local clustering coefficient and transitivity values
for this model are to the left of the bidisperse force-modified RGG model.

For mean shortest-path length in the LCC, we observe that, in contrast to our observations in 2D, the distribution of the bidisperse proximity-modified RGG is to the left of the bidisperse
force-modified RGG. Moreover, the distributions of the monodisperse and bidisperse force-modified
RGGs and the monodisperse proximity-modified RGG overlap more in 3D than in 2D. We also observe these differences between 2D and 3D in the distributions
of geodesic edge betweenness. The distributions of subgraph centrality and communicability in 3D are qualitatively more similar to the ones in 2D than is the case for the other
network diagnostics.

%
%
%


\section{Conclusions and discussion}
\label{sec:discussion}
We described and computed several common network diagnostics for a variety of 2D and 3D models of granular networks, ranging from unphysical ones with overlapping particles
to physical ones that are free (or at least mostly free) of overlaps. We studied the convergence properties
of the diagnostics to identify a reasonable number of network realizations for the computation of the
diagnostic distributions

We examined the ability of the various network diagnostics to distinguish between physical and
unphysical models of granular networks in 2D and 3D. Our results suggest that mean
local clustering coefficient and transitivity, which are related measures of triadic closure, are appropriate diagnostics to distinguish between the physical and unphysical models. The power of these two measures
lies in the fact that their distributions for the different models are readily distinguishable from each other.
In some cases, the diagnostics distributions of the 2D and 3D networks
appear to satisfy markedly different statistical distributions, a phenomenon that we also expect to observe
in comparisons of 2D and 3D granular networks from experiments.

Our results advance spatial random-graph models of granular networks and illustrate that simple
modifications of random geometric graphs that incorporate a minimal amount of physics are useful
models for gaining insights into granular and particulate networks. In future studies, it will be useful to
investigate the effects of shape variations, such as by examining the effects of different rectangular box
geometries~\cite{estrada2015random} and irregular boundaries~\cite{dettmann2015more} on network diagnostics in spatial random-graph models.


\acknowledgements

We thank Karen E.~Daniels for providing the experimental granular network data~\cite{bassett2012influence}. We also thank David Kammer for the use of his cluster computing infrastructure and Kornel Kovacs for the cluster maintenance. 


%
%
%
\newpage
\appendix
%
%
%


\section{Force-modified configurations}
\label{sec:overlap}
\begin{figure}
\begin{minipage}{0.41\textwidth}
\centering
\includegraphics[width=\textwidth]{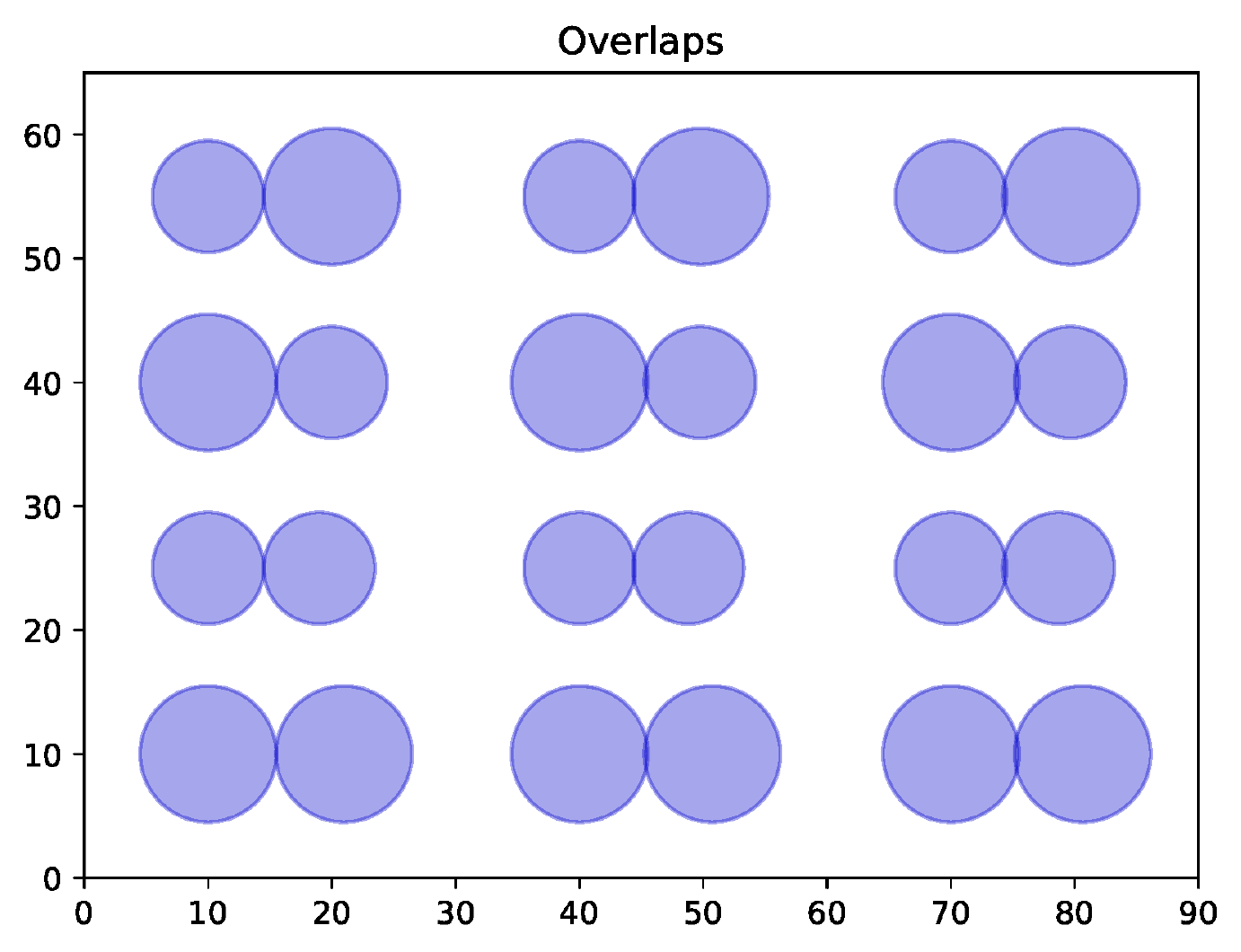}
\end{minipage}
\caption{\textbf{Overlapping particles.} The left column shows particles with an overlap of $r_{k}/200$, where $k$ indicates the particle on the left. In the middle column, the overlap is $r_{k}/20$; and in the right column, it is $r_{k}/15$.} 
\label{fig:overlaps}
\end{figure}
To obtain the potential-energy thresholds $V^\ast$ (below which we no longer update particle positions) in Sec.~\ref{subsec:force_mod_RGG}, we consider the first term of Eq.~\ref{eq:potential} and set the distance between two neighboring particles, $k$ and $l$, to be $r_{k}/200$ (which depends only on one of the particles). In a monodisperse packing, we use $r_k=r$, whereas $r_k \in {r_1 , r_2 }$ in bidisperse packings. For simplicity, we consider the case in which every particle has $6$ neighbors. In a monodisperse packing, we use $1022$ particles of radius $r=4.5~\text{mm}$ to determine a value of the potential-energy threshold. In a bidisperse packing, we consider a system with $561$ particles of radius $r_{1}=4.5~\text{mm}$ and $561$ with radius $r_{2}=5.5~\text{mm}$. 

Supposing that the radii are unitless, the potential-energy thresholds are
\begin{equation}
	\begin{aligned}
V_{\text{mono}}^\ast &= \sum_{k}\sum_{l \neq k} \left[\frac{1}{2}(r_{k} + r_{l}) - \frac{1}{2} |\mathbf{x}_{k} - \mathbf{x}_{l}|\right]_+^{\beta+1} \\
&= \sum_{k} 6 \times \left(\frac{r_{k}}{200}\right)^{\beta + 1} = 1122 \times 6 \times \left( \frac{4.5}{200}\right)^{1.5} \\
&\approx 22.7
	\end{aligned}
\end{equation}
for the monodisperse packing and
\begin{equation}
	\begin{aligned}
V_{\text{bi}}^\ast &= \sum_{k}\sum_{l \neq k} \left[\frac{1}{2}(r_{k} + r_{l}) - \frac{1}{2} |\mathbf{x}_{k} - \mathbf{x}_{l}|\right]_+^{\beta+1} \\
&= \sum_{k} 6 \times \left(\frac{r_{k}}{200}\right)^{\beta + 1} = 561 \times 6 \times \left( \frac{4.5}{200}\right)^{1.5} + 561 \times 6 \times \left( \frac{5.5}{200}\right)^{1.5} \\
&\approx 26.7 
	\end{aligned}
\end{equation}
for the bidisperse packing. In both cases, we set $\beta=1/2$, which is the value that we used for our computations in Sec.~\ref{sec:granular_models}. These two values of the potential-energy threshold provide an estimate to assess the quality of network configurations that we generated using the force-modified RGG model. For 3D models, we take the number of neighbors of a particle to be $12$. Therefore, the thresholds in 3D are $V^\ast_{\text{mono}} \approx 45.4$ and $V^\ast_{\text{bi}}\approx 53.4$.

We also examine the maximum overlap of our configurations. Based on visual inspection of Fig.~\ref{fig:overlaps}, we conclude that an overlap of $r_{k}/200$ is sufficiently small. However, overlap distances of $r_{k}/20$ and $r_{k}/15$ are visible.


\section{Diagnostic values}
\label{sec:diagnostic_values}
We present the numerical results for the diagnostics of the various models in Tables~\ref{tab:diag_exp_hex_RGG} and \ref{tab:diag_prox_force}.
\begin{table}[H]
\centering
\resizebox{0.95\columnwidth}{!}{
\begin{tabular}{@{}lcccc@{}}
\toprule
                                                  & \multicolumn{2}{c}{Experimental data} & \multicolumn{2}{c}{Modified lattice} \\ \midrule
\hline
Diagnostic                                        & Mean        & Standard deviation     & Mean       & Standard deviation    \\
\hline
Edge density & 0.0038 & $<10^{-4}$ & 0.0042 & $<10^{-4}$ \\
Geodesic node betweenness & 0.0163 & 0.0002 & 0.0168 & 0.0001 \\
Geodesic edge betweenness & 0.0080 & 0.0002 & 0.0083 &$<10^{-4}$ \\
Clustering coefficient & 0.2613 & 0.0100 & 0.3019 & 0.0068 \\
Transitivity & 0.2714 & 0.0070 & 0.2994 & 0.0042 \\
Degree assortativity & 0.1377 & 0.0236 & 0.0648 & 0.0262 \\
Global efficiency & 0.0772 & 0.0005 & 0.0812 & 0.0003 \\
Local efficiency & 0.3301 & 0.0156 & 0.3886 & 0.0089 \\
Mean shortest-path length in the LCC & 19.1634 & 0.1069 & 18.1703 & 0.0585 \\
Weighted mean shortest-path length & 19.1483 & 0.1046 & 18.1456 & 0.0680 \\
Maximized modularity & 0.8712 & 0.0045 & 0.8487 & 0.0028 \\
Subgraph centrality & 7.6236 & 0.2958 & 8.0579 & 0.0678 \\
Communicability & 0.0790 & 0.0046 & 0.0937 & 0.0013 \\ \bottomrule
\hline
\end{tabular}
}
\resizebox{0.95\columnwidth}{!}{
\begin{tabular}{@{}lcccc@{}}
\toprule
                                                  & \multicolumn{2}{c}{RGG (mono)} & \multicolumn{2}{c}{RGG (bi)} \\ \midrule
\hline
Diagnostic                                        & Mean        & Standard deviation     & Mean       & Standard deviation    \\
\hline
Edge density & 0.0038 & $<10^{-4}$ & 0.0038 & $<10^{-4}$ \\
Geodesic node betweenness & 0.0057 & 0.0046 & 0.0059 & 0.0046 \\
Geodesic edge betweenness & 0.0028 & 0.0022 & 0.0029 & 0.0022 \\
Clustering coefficient & 0.5485 & 0.0126 & 0.5480 & 0.0121 \\
Transitivity & 0.5921 & 0.0119 & 0.5885 & 0.0113 \\
Degree assortativity & 0.5839 & 0.0398 & 0.5581 & 0.0409 \\
Global efficiency & 0.0225 & 0.0045 & 0.0232 & 0.0047 \\
Local efficiency & 0.6314 & 0.0128 & 0.6304 & 0.0127 \\
Mean shortest-path length in the LCC & 24.6106 & 6.4311 & 24.4985 & 6.2803 \\
Weighted mean shortest-path length & 15.9323 & 5.2432 & 15.9851 & 5.1460 \\
Maximized modularity & 0.9528 & 0.0032 & 0.9521 & 0.0033 \\
Subgraph centrality & 35.0580 & 29.4708 & 34.9849 & 19.7010 \\
Communicability & 0.4376 & 0.6047 & 0.4336 & 0.3419 \\ \bottomrule
\hline
\end{tabular}
}
\caption{\textbf{Network diagnostics of the experimental data, the two-dimensional modified-lattice model,
and the two-dimensional RGGs.} We average the results over 17 realizations for the experimental data
and over 1023 realizations for the random-graph models.}
\label{tab:diag_exp_hex_RGG}
\end{table}
\begin{table}[H]
\centering
\resizebox{\columnwidth}{!}{
\begin{tabular}{@{}lcccc@{}}
\toprule
												  & \multicolumn{2}{c}{Proximity-modified RGG (mono)} & \multicolumn{2}{c}{Proximity-modified RGG (bi)} \\ \midrule
\hline
Diagnostic                                        & Mean               & Standard deviation             & Mean              & Standard deviation            \\
\hline
Edge density &0.0038 & $<10^{-4}$ & 0.0038 & $<10^{-4}$ \\
Geodesic node betweenness & 0.0171 & 0.0002 & 0.0166 & 0.0002 \\
Geodesic edge betweenness & 0.0085 & 0.0001 & 0.0082 & 0.0001 \\
Clustering coefficient & 0.3540 & 0.0070 & 0.3356 & 0.0066 \\
Transitivity & 0.3584 & 0.0044 & 0.3389 & 0.0040 \\
Degree assortativity & 0.2799 & 0.0270 & 0.2179 & 0.0268 \\
Global efficiency & 0.0734 & 0.0006 & 0.0754 & 0.0006 \\
Local efficiency & 0.4663 & 0.0100 & 0.4430 & 0.0099 \\
Mean shortest-path length in the LCC & 20.1977 & 0.1987 & 19.5844 & 0.1852 \\
Weighted mean shortest-path length & 20.1808 & 0.2012 & 19.5716 & 0.1858 \\
Maximized modularity & 0.8754 & 0.0031 & 0.8670 & 0.0032 \\
Subgraph centrality & 8.6839 & 0.2133 & 8.5141 & 0.2257 \\
Communicability & 0.0874 & 0.0036 & 0.0879 & 0.0040 \\ \\ \bottomrule
\hline
\end{tabular}
}
\resizebox{\columnwidth}{!}{
\begin{tabular}{@{}lcccc@{}}
\toprule
                                                  & \multicolumn{2}{c}{Force-modified RGG (mono)} & \multicolumn{2}{c}{Force-modified RGG (bi)} \\ \midrule
\hline
Diagnostic                                        & Mean               & Standard deviation             & Mean              & Standard deviation            \\
\hline
Edge density & 0.0038 & $<10^{-4}$ & 0.0038 & $<10^{-4}$ \\
Geodesic node betweenness & 0.0175 & 0.0004 & 0.0162 & 0.0001 \\
Geodesic edge betweenness & 0.0086 & 0.0002 & 0.0081 & 0.0001 \\
Clustering coefficient & 0.3106 & 0.0076 & 0.2238 & 0.0089 \\
Transitivity & 0.3311 & 0.0050 & 0.2463 & 0.0080 \\
Degree assortativity & 0.4404 & 0.0357 & 0.1952 & 0.0370 \\
Global efficiency & 0.0717 & 0.0010 & 0.0774 & 0.0004 \\
Local efficiency & 0.4172 & 0.0123 & 0.2789 & 0.0137 \\
Mean shortest-path length in the LCC & 20.7794 & 0.3931 & 19.1593 & 0.0938 \\
Weighted mean shortest-path length & 20.6838 & 0.4227 & 19.1463 & 0.0968 \\
Maximized modularity & 0.8659 & 0.0043 & 0.8410 & 0.0023 \\
Subgraph centrality & 8.6134 & 0.2818 & 7.3340 & 0.2270 \\
Communicability & 0.0934 & 0.0055 & 0.0760 & 0.0037 \\ \bottomrule
\hline
\end{tabular}
}
\caption{\textbf{Network diagnostics of the 2D proximity-modified RGG and the 2D force-modified RGG.} We average the results over $1023$ realizations.}
\label{tab:diag_prox_force}
\end{table}
%
%


\section{Additional diagnostic distributions}
\label{sec:distributions2}
\begin{figure}[htpb]
\centering
\includegraphics[width=0.95\textwidth]{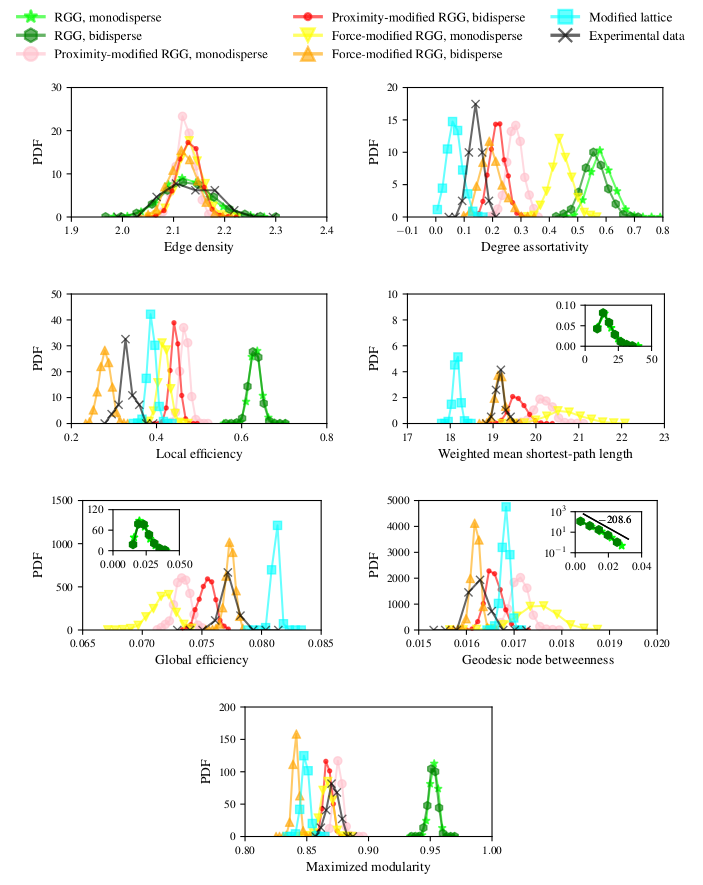}
\caption{\textbf{Distributions of the remaining diagnostics in the 2D models and experimental data.} We show the distributions of edge density, degree assortativity, local efficiency, weighted mean shortest-path length, global efficiency, geodesic node
betweenness and maximized modularity for the two-dimensional models and the experimental data. We determine the distributions
from 1023 network realizations for each model and for 17 realizations for the experimental data. In the plot of the edge-density
distribution, we do not show the distribution of the modified-lattice model, because the edge density for this model always has
exactly the same value. The black line in the inset of the plot of geodesic node betweenness is a guide to the eye that we obtain from
a power-law fit (using the method of least squares) of the distributions for standard bidisperse RGGs. (These insets also include
the distributions for the standard monodisperse RGGs, but the fits use data only from the bidisperse RGGs.) The corresponding
number indicates the slope of this line.
} 
\label{fig:diag_distribution_other}
\end{figure}

\begin{figure}[htpb]
\includegraphics[width=0.95\textwidth]{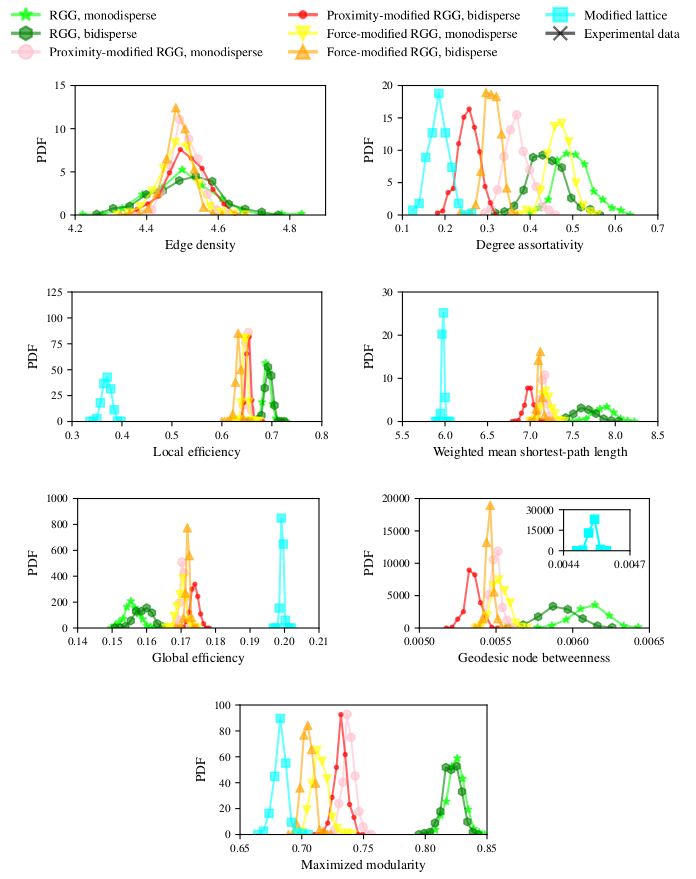}
\caption{\textbf{Distributions of the remaining diagnostics in the 3D models.} We show the distributions of edge density,
degree assortativity, local efficiency, weighted mean shortest-path distance, global efficiency, geodesic node betweenness and
maximized modularity for the three-dimensional models. We determine the distributions from 256 network realizations for each
model. In the plot of the edge-density distribution, we do not show the distribution of the modified-lattice model, because the
edge density for this model always has exactly the same value. For geodesic node betweenness, we show the distribution for the
modified-lattice model in the inset.
}
\label{fig:3D_diag_distribution_other}
\end{figure}

In Figs.~\ref{fig:diag_distribution_other} and \ref{fig:3D_diag_distribution_other}, we show the distributions of the diagnostics that we did not show in Sec.~\ref{sec:results}.

\end{document}